\documentclass[aps,preprint,tightenlines,showpacs,nofootinbib]{revtex4}

\usepackage{graphicx}
\usepackage{amsmath}
\usepackage{bm}

\newcommand{\beq}{\begin{equation}}
\newcommand{\eeq}{\end{equation}}
\newcommand{\bea}{\begin{eqnarray}}
\newcommand{\eea}{\end{eqnarray}}

\renewcommand{\vec}[1]{{\bm #1}}

\begin{document}
\title{On the limit cycle for the $1/r^2$ potential in momentum space}

\author{H.-W. Hammer}%\email{hammer@phys.washington.edu}
\affiliation{Institute for Nuclear Theory, University of Washington, Seattle,
        WA\ 98195, USA}

\author{Brian G.\ Swingle}

\affiliation{School of Physics,
Georgia Institute of Technology, Atlanta, GA 30332, USA}
\affiliation{Institute for Nuclear Theory, University of Washington, Seattle,
        WA\ 98195, USA}

\date{\today}

\begin{abstract}
The renormalization of the attractive $1/r^2$ potential has recently been 
studied using a variety of regulators. In particular, it was shown
that renormalization with a square well in position space
allows multiple solutions for the depth of the square well,
including, but not requiring a renormalization group limit cycle.
Here, we consider the  renormalization of the
$1/r^2$ potential in momentum space. We regulate the problem with a 
momentum cutoff and absorb the cutoff dependence using a momentum-independent
counterterm potential.
The strength of this counterterm is uniquely determined and runs 
on a limit cycle. We also calculate the bound state spectrum and scattering
observables, emphasizing the manifestation of the limit cycle in these
observables.
\end{abstract}


\maketitle

\section{Introduction}
The renormalization of the quantum mechanical $1/r^2$ potential
has attracted some interest recently because this
singular potential is one of the simplest examples displaying
discrete scale invariance and a renormalization group limit cycle
\cite{Beane:2000wh,Bawin:2003dm,Braaten:2004pg,Barford:2004fz}.

The renormalization group (RG) is an important tool in
many areas of modern physics \cite{Wilson:dy}. Its applications
range from critical phenomena in condensed matter physics to the
renormalization of quantum field theories in nuclear and particle
physics. The solutions of the RG can have many different topologies. 
Most applications, however, involve a renormalization group flow
towards a (scale invariant) fixed point. A prominent example is
Quantum Chromodynamics (QCD) the theory of the strong
interactions. QCD has a single coupling constant
$\alpha_s(\Lambda)$ with an asymptotically free ultraviolet fixed
point: $\alpha_s(\Lambda) \to 0$ as the momentum cutoff $\Lambda$ is 
taken to infinity \cite{Gross:1973id,Politzer:fx}. 

The possibility of an RG limit cycle was already pointed out by Wilson in 
1971 \cite{Wilson:1970ag}. A limit cycle is a closed curve in the space
of coupling constants that is invariant under the RG flow. The RG
flow completes a cycle around the curve every time the cutoff is
changed by a multiplicative factor $\lambda_0$. This number $\lambda_0$
is called the {\it preferred scaling factor}. A limit cycle
requires at least two coupling constants coupled by the RG
equations if the couplings are continuous \cite{Wilson:1970ag}.
With a single coupling constant, a limit cycle can only occur if 
the coupling has discontinuities. A necessary
condition for a limit cycle is invariance under discrete scale
transformations: $x\to \lambda_0^n x$, where $n$ is an integer. This
discrete scaling symmetry is reflected in log-periodic behavior of
physical observables. It is interesting to observe that discrete
scale invariance also arises in a variety of other contexts such as
turbulence, sandpiles, earthquakes, and financial crashes
\cite{Sor97}. We also note that because of the $c$-theorem in two spatial 
dimensions \cite{Zamolodchikov:gt}, limit cycles are expected to occur only 
in $D \geq 3$ dimensions.\footnote{However, see Ref.~\cite{Leclair:2003xj}
for an apparent counter example.}

One important example of a limit cycle that was identified long ago from
its manifestation in the bound state spectrum, is the three-body problem
with large scattering length $a$ \cite{Albe-81}.
In the limit $a \to \pm \infty$, there is an accumulation of
3-body bound states near threshold with binding energies differing by
multiplicative factors of $\lambda_0^2 \simeq 515.03$ \cite{Efimov70}.
This phenomenon -- the Efimov effect -- can be understood
in terms of a renormalization group limit cycle with
discrete scaling factor $\lambda_0 \simeq 22.7$.

The three-body problem with large $a$
is intimately connected to the quantum-mechanical $1/r^2$ potential.
In the ultraviolet limit of large three-momenta, the three-body problem 
can be mapped into a one-dimensional Schr\"odinger equation with a
$1/\rho^2$ potential where $\rho^2=r_{12}^2+r_{13}^2+r_{23}^2$ is the
hyperradius of the three particles
($r_{ij}\equiv |\vec{r}_i-\vec{r}_j|$)  \cite{Efi71}.
As a consequence, the renormalization of both problems is similar.

In the effective field theory (EFT) formulation of the three-body
problem with large scattering length,
the limit cycle is evident in the RG evolution of a contact
three-body interaction which is required for renormalization
at leading order in the EFT expansion \cite{Bedaque:1998kg,BHvK99b}.
The limit cycle behavior has important consequences in nuclear and atomic
three-body systems \cite{BrH04}. For example,
it has been conjectured that QCD has an infrared RG limit cycle
at special values of the quark masses \cite{Braaten:2003dw}.
Limit cycles have recently also been realized in discrete Hamiltonian
models \cite{Glazek:2002hq,Glazek:2004}, superconductivity 
\cite{LeClair:2002ux,Anfoss05}, quantum field theory models 
\cite{Leclair:2003xj,Bernard:2001sc}, and S-matrix models 
\cite{LeClair:2003hj,LeClair:2004ps}.

Since discrete scale invariance occurs in a wide variety
of complex systems \cite{Sor97}, RG limit cycles may play a more important
role in physics than previously realized. In this context, 
the study of the attractive $1/r^2$ potential sheds some light on the 
general properties of limit cycles and the mechanism of their emergence.
Furthermore, many properties of the EFT approach to the three-body system 
with large $a$ can be illustrated using this simpler problem.

The quantum-mechanical $1/r^2$ potential
has a long and venerable history \cite{Case:1950,FLS71}.
Here, we are most interested in the renormalization group
aspects of this problem.\footnote{For more references to 
previous work on $1/r^2$ potential
emphasizing other aspects of the problem, see Ref.~\cite{Cam00}.}
The renormalization of the $1/r^2$ potential has been studied 
in position space within the renormalization group
framework by two different groups using a spherical square-well
regulator potential \cite{Beane:2000wh,Bawin:2003dm}.
Beane et al. found that there are infinitely many
choices for the strength of the square-well regulator,
including continuous functions of the short-distance
cutoff $R$ as well as a log-periodic function of $R$
with discontinuities corresponding to an RG limit cycle \cite{Beane:2000wh}.
Bawin and Coon obtained a closed-form solution for the coupling constant 
that is log-periodic, which suggests that the choice with the RG limit cycle 
is in some sense natural \cite{Bawin:2003dm}.
An alternative regularization with a delta-shell potential
was considered by Braaten and Phillips \cite{Braaten:2004pg}. 
They have argued that a
limit cycle is the most natural choice for the renormalization group
behavior of the regulator potential and shown that a limit
cycle is unavoidable for the renormalization with a delta-shell 
potential. In Ref.~\cite{MuH04}, the limit cycle for the $1/r^2$
potential has been studied using RG flow equations.
Most recently, Barford and Birse have applied a distorted wave 
RG to scattering by an inverse square potential and three-body systems
with large scattering length $a$ \cite{Barford:2004fz}.

In this paper, we consider the renormalization of the $1/r^2$ potential
in momentum space. Our approach is very similar
to the EFT treatment of the three-body system with large $a$
\cite{Bedaque:1998kg,BHvK99b}. We add a momentum-independent
counterterm potential and determine its renormalization group 
behavior from invariance of the low-energy observables under
variations of the ultraviolet cutoff $\Lambda$.

\section{The $1/r^2$ Potential in Momentum Space}
\label{sec:mom}

We consider the attractive inverse square potential
\beq
V(r) = \frac{\hbar^2}{m}\frac{c}{r^2}\,,
\qquad\mbox{with}\quad r\equiv|\vec{r}|\,,\quad
c\equiv -\frac{1}{4}-\nu^2\,,
\label{eq:defpot}
\eeq
and $\nu > 0$ a positive real parameter.
This potential has the same scaling behavior as the kinetic energy 
operator and, consequently, is scale invariant at the classical level.
In the following, we set the particle
mass and Planck's constant $m=\hbar=1$ for convenience.
For values of $c\geq -\frac{1}{4}$, the potential is well-behaved and
the corresponding Schr\"odinger equation has a unique solution, 
see Ref.~\cite{FLS71}.
We are interested in the case $c< -\frac{1}{4}$ which corresponds
to real values of $\nu$ in (\ref{eq:defpot}).
In this case, the $1/r^2$ potential is singular and the
usual boundary conditions for the Schr\"{o}dinger equation do not
lead to a unique solution. 
Mathematically, this problem occurs because the $1/r^2$ 
potential is not self-adjoint. It can be cured by defining
a self-adjoint extension of the potential which leads to a unique
solution of the Schr\"{o}dinger equation 
(see, e.g., Ref.~\cite{Bawin:2003dm} and references therein).

We use a physically more intuitive way of dealing with the
non-uniqueness and interpret the $1/r^2$ potential as an effective
theory \cite{Kaplan:1995uv,Lepage:1997cs}. This approach can
easily be extended to field-theoretical problems.
In the effective theory framework,
the singular behavior at $r=0$ is not physical since the potential
is modified at short distances by physics not included in the effective
theory. The singular behavior of the potential is regulated using
some form of short-distance cutoff.
The low-energy observables can be made independent of the regulator by
including a short-distance counterterm in the effective theory
that captures the effect of the unknown short-distance physics on
low-energy observables. 

First we require the $1/r^2$ potential in momentum space.
We can calculate the Fourier transform of the potential using
dimensional regularization. We define the Fourier integrals in $D$
dimensions and evaluate the integrals for values of $D$
for which they are convergent. In the end, we analytically
continue back to $D=3$. Using the definitions 
\beq 
V(q)=\lim_{D\to 3}\int d^D \vec{r}\, e^{i\vec{q}\cdot\vec{r}}\, V(r)\qquad
\mbox{and} \quad 
V(r)=\lim_{D\to 3}\int \frac{d^D \vec{q}}{(2\pi)^3}\, 
e^{-i\vec{q}\cdot\vec{r}} \,V(q)\,, 
\eeq 
we obtain the expression 
\beq 
V(q)=\frac{2\pi^2 c}{q}\,
\label{eq:FT1oR2} 
\eeq 
for the momentum space representation of the $1/r^2$ potential 
(\ref{eq:defpot}). Since the potential is local, its Fourier transform 
depends only on the momentum transfer $q$.

\section{Renormalization}
\label{sec:renorm}

We are now in the position to calculate low-energy observables
using the Lippmann-Schwinger equation.
We start with the bare potential $V(q)$ from Eq.~(\ref{eq:FT1oR2})
and illustrate the non-uniqueness
of the resulting Lippmann-Schwinger equation. Then, we demonstrate
how to renormalize the equation using a momentum cutoff and
a counter term $\delta V$.

\begin{figure}[ht]
\centerline{\includegraphics*[width=10cm,angle=0]{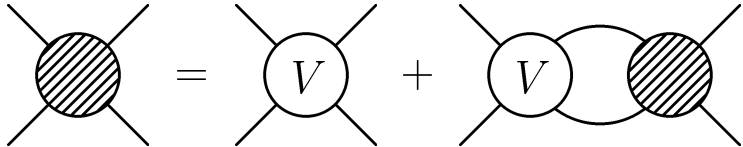}}
\caption{Lippmann-Schwinger equation for a two-body potential $V$.}
\label{fig:LS}
\end{figure}

The Lippmann-Schwinger equation for two particles interacting
in via $V(q)$ from Eq.~(\ref{eq:FT1oR2}) in their center-of-mass frame
takes the form
\beq
t_E (\vec{p},\vec{p}') = \frac{2\pi^2 c}{|\vec{p}-\vec{p}'|}
+\int \frac{d^3 q}{(2\pi)^3} \frac{2\pi^2 c}{|\vec{p}-\vec{q}|}
   \frac{t_E(\vec{q} ,\vec{p}')}{E-q^2+i\epsilon}\,,
\label{eq:LSeqraw}
\eeq
where $E$ is the total energy and
$\vec{p}$ ($\vec{p}'$) are the relative momenta of the incoming (outgoing)
particles, respectively. A pictorial representation of this equation is
given in Fig.~\ref{fig:LS}.
%%%%%%%%%%%%%%%%%%%%%%%%%%%%%%%%%%%%%%%%%%%%%%%%%%%%%%%%%%%%%%%%%%%%%%%%
%%%%%%%%%%%%%%%%%%%%%%%%%%%%%%%%%%%%%%%%%%%%%%%%%%%%%%%%%%%%%%%%%%%%%%%%
We are only interested in the S-waves. In the higher partial 
waves the singular behavior of the potential is 
screened by the angular momentum barrier. Projecting onto S-waves by
integrating the equation over the relative angle between $\vec{p}$ and
$\vec{p}'$: $\frac{1}{2}\, \int d\cos\theta_{\vec{p}\vec{p}'}$,
we obtain the integral equation
\beq
t_E (p,p')=2\pi^2 c f(p,p') + c\int_0^\infty \frac{dq\,q^2}{E-q^2+i\epsilon}
\,f(p,q)\,t_E (q,p')\,,
\label{eq:LSeq}
\eeq
where
\beq
f(p,q)=\frac{\theta(p-q)}{p}+\frac{\theta(q-p)}{q}\,.
\eeq

The physical observables are the bound state spectrum
and the scattering phase shifts $\delta(k)$. Note, however, that
$\delta(k)$ for the $1/r^2$ potential
is only defined for $k>0$ \cite{FLS71}.
The phase shifts are determined by the
solution to Eq.~(\ref{eq:LSeq}) evaluated at the on-shell point $E=k^2$,
$k=p'=p$ via
\beq
T(k)=\frac{1}{k\cot\delta(k)-ik}=-\frac{1}{4\pi} t_{k^2}(k,k)\,.
\eeq
Since $p'$ appears only as a parameter in Eq.~(\ref{eq:LSeq}),
we can set $p'=k$ to simplify the equation.
The binding energies are given by those values of $E<0$ for which
the homogeneous version of Eq.~(\ref{eq:LSeq}) has a solution.
For the bound state equation the dependence of the solution $\phi_E(p)$
on $p'$ disappears altogether.

It is well-known that Eq.~(\ref{eq:LSeq}) does not have a unique solution 
since the $1/r^2$ potential for real $\nu$ is singular \cite{FLS71}.
This can easily be seen by considering the zero-energy
bound state solution $\phi_0 (p)$:
\beq
\phi_0 (p)=-c\left[\int_0^p \frac{dq}{p}\, \phi_0(q)
  +\int_p^\infty \frac{dq}{q}\,\phi_0 (q)\right]\,.
\label{eq:LSasy}
\eeq
Since this equation is scale invariant, its solution is a power law.
Inserting the standard ansatz $\phi_0 (p)=p^s$ into Eq.~(\ref{eq:LSasy}),
we obtain the consistency condition
\beq
s^2+s-c=0\,, \qquad -1 < {\rm Re}\,s < 0\,,
\eeq
which has the two solutions
\beq
s_\pm = -\frac{1}{2} \pm i\nu\,.
\eeq
Neither of the two solutions can be excluded on physical grounds.
They are both oscillatory functions of $\ln p$ and vanish as 
$p\to \infty$. Therefore the most general solution to Eq.~(\ref{eq:LSasy}) 
can be written as
\beq
\phi_0 (p)={\cal N}\, p^{-1/2} \left(p^{i\nu}e^{i\alpha}+
   p^{-i\nu}e^{-i\alpha}\right)\,,
\label{eq:2sol}
\eeq
where the relative phase $\alpha$ is a free parameter. 
The value of $\alpha$ is not determined by the $1/r^2$ 
potential and has to be taken from elsewhere. It is exactly this phase
$\alpha$ which is fixed by self-adjoint extensions of the
potential \cite{Bawin:2003dm}.

In our case, this is conveniently done using renormalization theory.
We regularize the Lippmann-Schwinger equation by applying a momentum cutoff
$\Lambda$ and include a momentum-independent counter term
$\delta V(\Lambda)$ in the potential.
The precise form of the cutoff, for example Gaussian cutoff or sharp
cutoff, is not important, but we use a sharp cutoff for simplicity.
Making the replacement
\beq
V(q) = \frac{2\pi^2 c}{q} \quad \Longrightarrow \quad
V(q)+\delta V (\Lambda)= 2\pi^2 c\,\left(\frac{1}{q} + 
\frac{H(\Lambda)}{\Lambda}\right)\,,
\eeq
the Lippmann-Schwinger equation (\ref{eq:LSeq}) becomes
\beq
t_E (p,k)=2\pi^2 c \left[f(p,k)+\frac{H(\Lambda)}{\Lambda}\right]
+ c\int_0^\Lambda \frac{dq\,q^2}{E-q^2+i\epsilon}
\,\left[f(p,q)+\frac{H(\Lambda)}{\Lambda}\right]\,t_E (q,k)\,.
\label{eq:LSeqCT}
\eeq
The functional dependence of $H(\Lambda)$ can be determined
analytically from invariance
under the renormalization group. We demand that the relative phase
of the zero-energy bound solution of Eq.~(\ref{eq:LSeqCT})
remains unchanged under variations of the cutoff $\Lambda$.
As we demonstrate below, this is sufficient to keep all low-energy
observables independent of $\Lambda$. We find
\beq
H(\Lambda)=\frac{1-2\nu\tan(\nu\ln(\Lambda/\Lambda_*))}
         {1+2\nu\tan(\nu\ln(\Lambda/\Lambda_*))}\,,
\label{eq:Hdep}
\eeq
where $\Lambda_*$ is a free parameter that determines the relative
phase in (\ref{eq:2sol}): $\alpha=-\nu\ln\Lambda_*$. 
%%%%%%%%%%%%%%%%%%%%%%%%%%%%%%%%%%%%%%%%%%%%%%%%%%%%%%%%%%%%%%%%%%%%%%%%
\begin{figure}[ht]
\centerline{\includegraphics*[width=8cm,angle=0]{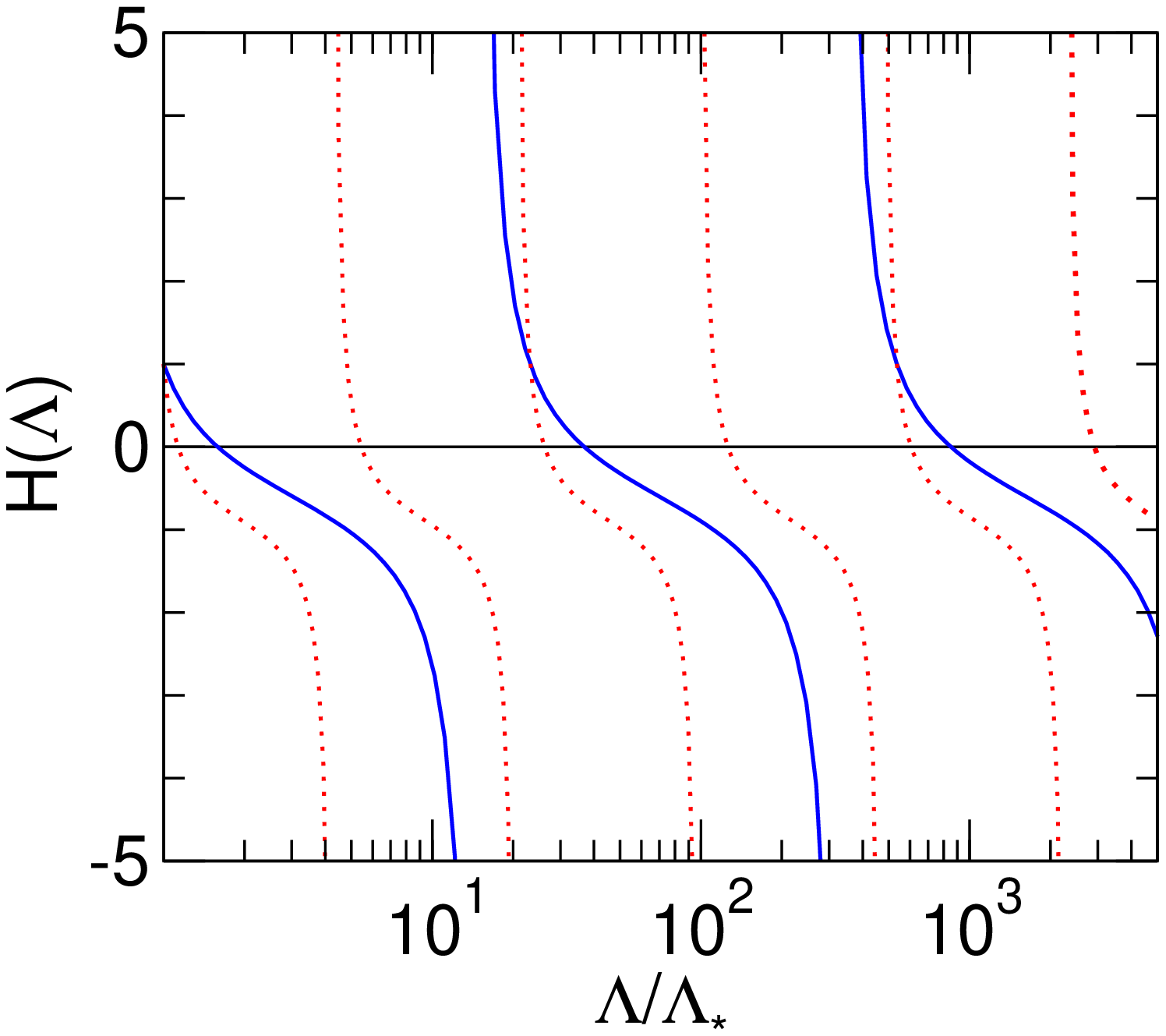}
\quad
\includegraphics*[width=8.3cm,angle=0]{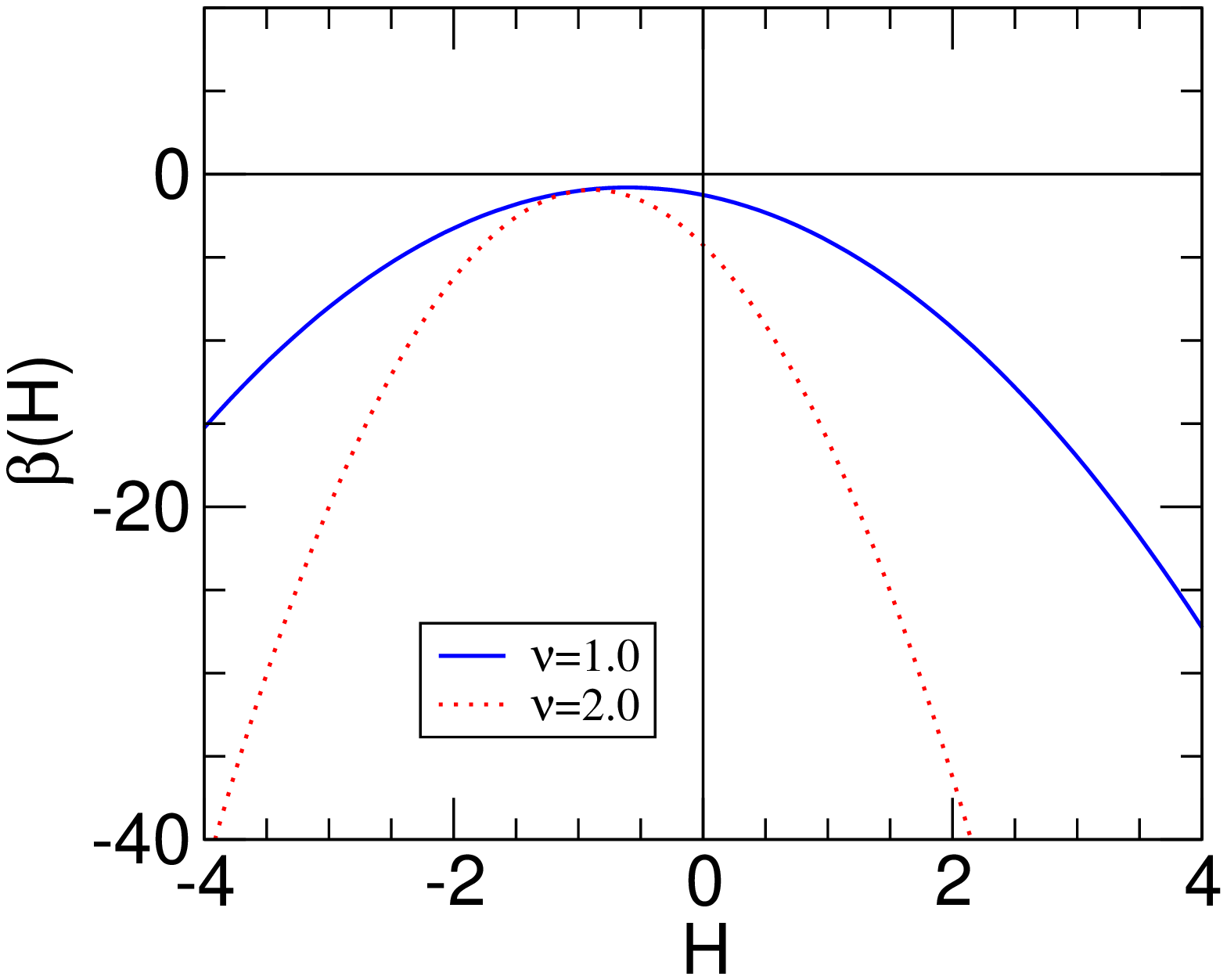}}
\caption{Left panel: The dimensionless coupling $H(\Lambda)$.
Right panel: The $\beta$ function $\beta(H)$. The solid (dotted)
line corresponds to $\nu=1$ ($\nu=2$).
}
\label{fig:HoL}
\end{figure}
%%%%%%%%%%%%%%%%%%%%%%%%%%%%%%%%%%%%%%%%%%%%%%%%%%%%%%%%%%%%%%%%%%%%%%%%
In order to fix the relative phase $\alpha$ in Eq.~(\ref{eq:2sol}),
we can either specify both the cutoff $\Lambda$ and the dimensionless 
coupling $H$ or, using Eq.~(\ref{eq:Hdep}), one dimensionful parameter: 
$\Lambda_*$. This parameter $\Lambda_*$ is generated by the iteration of
quantum corrections in solving the integral equation  (\ref{eq:LSeqCT}).
In essence, this is the phenomenon of dimensional transmutation
\cite{Wil99}.

In the left panel of 
Fig.~\ref{fig:HoL}, we show the functional dependence of $H$ on
$\Lambda$ for $\nu=1$ (solid line) and $\nu=2$ (dashed line).
It is evident that $H(\Lambda)$ runs on a limit cycle with a preferred
scaling factor $\lambda_0=\exp(\pi/\nu)$. If the cutoff is increased
or decreased by a factor $\lambda_0$, $H(\Lambda)$ returns to its
original value. In contrast to the square-well regularization in
configuration space \cite{Beane:2000wh}, it is not possible to
choose a continuous function for $H(\Lambda)$. The scale invariance
of the bare potential is not compatible with the regulator.
The renormalization breaks the full scale invariance of the $1/r^2$ 
potential (\ref{eq:defpot}) down to the discrete sub-group of scaling
transformations with the preferred scaling factor $\lambda_0$.
This is a simple example of a quantum-mechanical anomaly
\cite{Braaten:2004pg}.\footnote{For a
discussion of the $1/r^2$ potential and anomalies in conformal
quantum mechanics, see Ref.~\cite{Camblong:2003mz}.}

Note also that  $H(\Lambda)$ vanishes for a special set of cutoffs
\beq
\Lambda_n(\Lambda_*)=\Lambda_* \exp(n\pi/\nu)\,,
\label{eq:lambdaN}
\eeq
where $n$ is an integer.
We can therefore obtain a renormalized version of Eq.~(\ref{eq:LSeqCT})
that does not explicitly contain the counterterm by using
the discrete set of cutoffs from Eq.~(\ref{eq:lambdaN}).
The same trick can be used for the three-body problem with large
scattering length \cite{Hammer:2000nf}.

Further insight can be gained by deriving the RG equation 
for $H$. Using the explicit solution (\ref{eq:Hdep}), we find for the
$\beta$ function by differentiating:
\beq
\Lambda\frac{d}{d\Lambda} H \equiv \beta(H) =
-\frac{1}{4}(1-H)^2-\nu^2 (1+H)^2\,.
\label{eq:beta}
\eeq
This expression for $\beta(H)$ is shown in the right panel of 
Fig.~\ref{fig:HoL} for the two cases $\nu=1$ (solid line) and 
$\nu=2$ (dotted line).
Since $\beta(H)$ is negative definite for $\nu>0$, there are no fixed point 
solutions for $H$. The $\beta$ function has only two complex roots,
$H_{\pm}=-(2\nu \pm i)/(2\nu \mp i)\,.$
It becomes maximal at $H_{max}=-(\nu^2-1/4)/(\nu^2+1/4)$ and takes the
value $\beta(H_{max}) = -\nu^2/(\nu^2+1/4)$. As $H$ approaches $\pm \infty$ 
the $\beta$ function approaches $-\infty$. In the limit $\nu\to 0$
($\nu \to \infty$), the limit cycle disappears and a fixed point
at $H=1$ ($H=-1$) emerges.

\section{Observables}
\label{sec:dshreg}

We now turn to the calculation of low-energy observables for the 
$1/r^2$ potential. We demonstrate explicitly that the scattering phases 
and the binding energies are independent of $\Lambda$ if $H(\Lambda)$
varies as given in Eq.~(\ref{eq:Hdep}). In particular, we discuss
the manifestation of the limit cycle in the observables.

\begin{figure}[ht]
\centerline{\includegraphics*[width=10cm,angle=0]{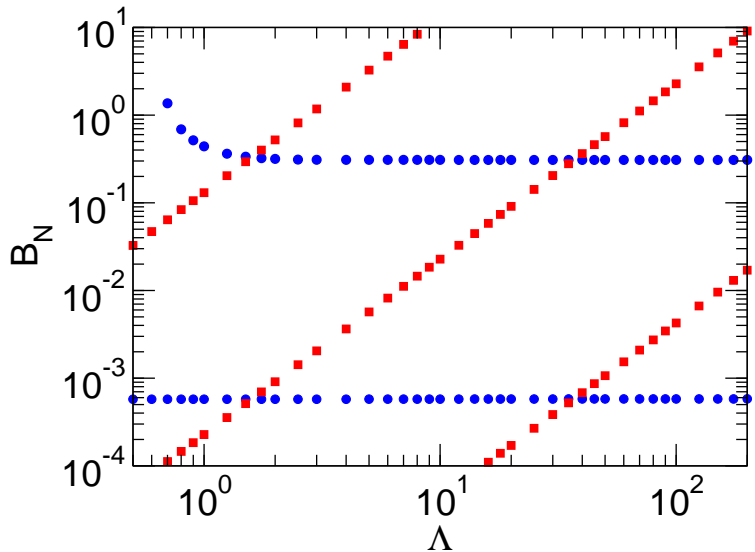}}
\caption{The bound state spectrum as a function of the cutoff for
$H(\Lambda)\equiv0$ (squares) and $H(\Lambda)$ given by
Eq.~(\ref{eq:Hdep}) (circles).
}
\label{fig:Bcut}
\end{figure}

First, we consider the bound state spectrum.
In Fig.~\ref{fig:Bcut}, we show a part of the spectrum as
%%%%%%%%%%%%%%%%%%%%%%%%%%%%%%%%%%%%%%%%%%%%%%%%%%%%%%%%%%%%%%%%%%%%%%%%
%%%%%%%%%%%%%%%%%%%%%%%%%%%%%%%%%%%%%%%%%%%%%%%%%%%%%%%%%%%%%%%%%%%%%%%%
a function of the cutoff $\Lambda$. The squares show the binding energies
for the unrenormalized case: $H(\Lambda)\equiv 0$. The binding energies
grow as the cutoff is increased and there is an accumulation point
at threshold. 
The spectrum is bounded from below and the
energy of the deepest state is of order $\Lambda^2$.
However, if $H(\Lambda)$ varies according to Eq.~(\ref{eq:Hdep}),
the binding energies are independent of $\Lambda$. The deepest
state ist still of order $\Lambda^2$, but a new deeply-bound state appears 
whenever the cutoff is increased by
a power of the preferred scaling factor $\lambda_0 =\exp(\pi/\nu)$. 
The shallow binding energies, however, are not affected. 
In particular, the accumulation point at threshold persists
since it is an infrared effect. This phenomenon is strongly
related to the Efimov effect in the three-body problem with large
scattering length where $\nu\approx 1.00624$ \cite{Efimov70}.
As a consequence, all bound states
within the range of the effective theory with binding energy
$B \ll \Lambda^2$ are renormalized.

It is straightforward to show that if $B$ is a binding energy, then
$\exp(2\pi/\nu)\,B$ is a solution as well. For dimensional reasons,
the dependence of the binding energies on the cutoff is
\beq
\ln B_N = c_1 -N \frac{2\pi}{\nu} +2\ln\Lambda\,,\qquad
N\geq 0\,,\quad c_1 = -2.07\pm 0.03\,,
\eeq
where we have arbitrarily labelled the deepest state with $N=0$.
The dependence of the renormalized energies on $\Lambda_*$ can be
obtained by using Eq.~(\ref{eq:lambdaN}).

\begin{figure}[ht]
\centerline{\includegraphics*[width=10cm,angle=0]{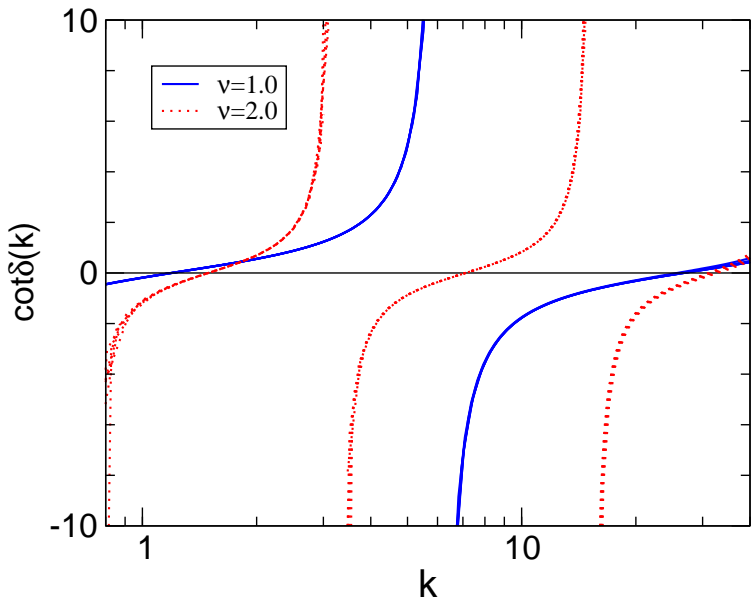}}
\caption{The quantity $\cot\delta$ as a function of $k$
for $\Lambda_*=1.0$
and $\nu=1$ (solid line) as well as $\nu=2$ (dotted line). Each curve
corresponds to various calculations with cutoffs in the range
$\Lambda=50\ldots 100$.
}
\label{fig:kcot}
\end{figure}

We now turn to the scattering observables. From now on, we only
consider the renormalized case.
In Fig.~\ref{fig:kcot}, we show $\cot\delta(k)$ 
as a function of the center-of-mass
momentum $k=\sqrt{E}$ for $\Lambda_*=1.0$
and two different values of $\nu$.
%%%%%%%%%%%%%%%%%%%%%%%%%%%%%%%%%%%%%%%%%%%%%%%%%%%%%%%%%%%%%%%%%%%%%%%%
%%%%%%%%%%%%%%%%%%%%%%%%%%%%%%%%%%%%%%%%%%%%%%%%%%%%%%%%%%%%%%%%%%%%%%%%
The solid line corresponds to $\nu=1$, while the dotted line
corresponds to $\nu=2$. Each curve
represents various calculations with cutoffs in the range
$\Lambda=50\ldots 100$. The spread of the lines shows the cutoff
dependence remaining after the counter term $\delta V$ is included.
Note that the renormalization from the previous section is exact only
at zero energy. For finite energy, there are corrections that are
suppressed by powers of $k/\Lambda$. 
For momenta $k \ll \Lambda$, they are small and the
phase shifts are practically independent of the cutoff $\Lambda$.

The discrete scaling symmetry requires that $\cot\delta(k)$ is a
periodic function of $\nu\ln(k/\Lambda_*)$. This is clearly
statisfied in Fig.~\ref{fig:kcot}. It follows that
the general form of the phase shifts in Fig.~\ref{fig:kcot} can
be written as
\beq
\delta (k)=\beta(\nu)-\nu\ln(k/\Lambda_*)\,,
\label{eq:delta}
\eeq
where $\beta$ is an angle that depends on $\nu$. From Eq.~(\ref{eq:delta})
and Fig.~\ref{fig:kcot}, it is evident that the phase shift is not 
defined in the limit $k\to 0$ \cite{FLS71}.

\begin{figure}[ht]
\centerline{\includegraphics*[width=10cm,angle=0]{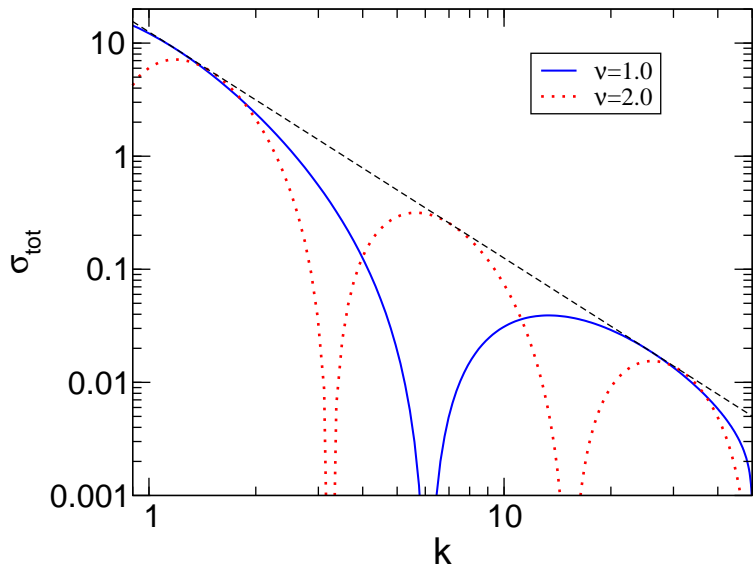}}
\caption{The total scattering cross section for $\Lambda_*=1.0$
and $\nu=1$ (solid line) as well as $\nu=2$ (dotted line).
The dashed line indicates the unitarity limit $4\pi/k^2$.
}
\label{fig:sigtot}
\end{figure}

Another scattering observable that can be obtained from the phase
shift is the total cross section for S-wave scattering 
\beq
\sigma_{tot}=\int d\Omega |T(k)|^2 =\frac{4\pi}{[k\cot\delta(k)]^2 +k^2}\,.
\eeq
In Fig.~\ref{fig:sigtot} we show $\sigma_{tot}$ as a function
of $k=\sqrt{E}$
%%%%%%%%%%%%%%%%%%%%%%%%%%%%%%%%%%%%%%%%%%%%%%%%%%%%%%%%%%%%%%%%%%%%%%%%
%%%%%%%%%%%%%%%%%%%%%%%%%%%%%%%%%%%%%%%%%%%%%%%%%%%%%%%%%%%%%%%%%%%%%%%%
for the same parameter values as in Fig.~\ref{fig:kcot}.
The solid line corresponds to $\nu=1$, while the dotted line
corresponds to $\nu=2$. The dashed line gives the unitarity limit
$\sigma_{uni}=4\pi/k^2$. The total cross section saturates the
unitarity limit exactly at those values of $k$ where $k\cot\delta (k)$
vanishes. It has zeros at the values of $k$ where $k\cot\delta (k)$
diverges. The dimensionless cross section $\sigma_{tot}/\sigma_{uni}$
is invariant under discrete scale transformations.

In summary, all low-energy observables display the discrete 
scaling symmetry. The limit cycle 
leads to log-periodic dependence of the observables on the 
counterterm parameter $\Lambda_*$. Furthermore, the limit cycle also 
becomes manifest in scattering observables via a log-periodic dependence 
on the energy. This is the case
even if the counterterm strength can be chosen as 
a continuous function of the cutoff (cf. Ref.~\cite{Beane:2000wh}).

\section{Summary \& Conclusions}
\label{sec:conc}

In this paper, we have studied the renormalization of the
singular $1/r^2$ potential in momentum space. 
This potential is one of the simplest physical systems with a 
RG limit cycle. Interpreting this potential as an effective theory that 
breaks down at short distances, we have renormalized the 
corresponding Lippmann-Schwinger equation for S-waves
by introducing a momentum cutoff $\Lambda$
and a momentum-independent
counter term potential $\delta V\propto H(\Lambda)/\Lambda$.

Demanding invariance of low-energy observables under variations 
of the cutoff $\Lambda$, we have shown that the  dimensionless function 
$H(\Lambda)$ runs on a limit cycle. The function $H(\Lambda)$ depends
on a dimensionful parameter $\Lambda_*$ that must be specified in 
addition to the strength of the $1/r^2$ potential $\nu$ in order 
to characterize a physical system uniquely.
Different values of $\Lambda_*$ correspond to
physical systems with the same long range behavior but
different short distance physics.

For our choice of regulator,
a continuous dependence of the counterterm on $\Lambda$ 
is not possible. This is in contrast to renormalization of the 
$1/r^2$ potential with a square well potential in position space where 
a continuous dependence can be chosen \cite{Beane:2000wh}.

The scale invariance of the bare $1/r^2$ potential is 
anomalous. It is not compatible
with the regularization of the singularity of this potential at $r=0$. 
The scale invariance is broken down to the discrete subgroup
of scaling transformations with the preferred scaling factor 
$\lambda_0=\exp(\pi/\nu)$ by the renormalization procedure.

The limit cycle of the counterterm potential becomes manifest 
in physical observables via the discrete scaling symmetry and
the corresponding log-periodic behavior. 
While the periodic dependence of the counterterm potential on the
cutoff $\Lambda$ appears to be dependent on the specific regulator, the 
discrete scaling symmetry is robust and regulator-independent.
We have illustrated the manifestation of this scaling symmetry in 
the log-periodic behavior of low-energy observables on the
energy by calculating the bound state spectrum,
the S-wave scattering phase shifts, and the total S-wave cross 
section. 

It will be interesting to see if the effect of limit cycles can be observed
in experiment. The three-body physics of cold alkali atoms near a
Feshbach resonance, where their interaction strength can be tuned 
by adjusting an external magnetic field, appears to be a promising avenue 
\cite{BrH04}.
In Ref.~\cite{Weber03}, the three-body recombination rate for
$^{133}$Cs atoms in the $|3,+3 \rangle$ hyperfine state was measured
in the interval 10 G $<B<$ 150 G which includes several Feshbach resonances.
Unfortunately, the log-periodic behavior could neither be observed nor 
could it be excluded. Improved experiments are called for to resolve this
question. Another indication of a limit cycle would be the observation
of the Efimov effect.
Recent experiments with Bose-Einstein condensates of $^{85}$Rb atoms near a 
Feshbach resonance have produced evidence for a condensate of diatomic 
molecules coexisting with the atom condensate \cite{Don02}.
It might be possible to create condensates of the triatomic molecules 
predicted by the Efimov effect coexisting with atom and dimer 
condensates \cite{Braaten:2002er}.
Finally, the attractive $1/r^2$ potential has been realized in experiment
by neutral atoms interacting with a charged wire \cite{Dens98}.

\acknowledgments

We thank U. van Kolck for a suggestion.
%We thank E.~Braaten for valuable discussions.
HWH was supported by the  Department of Energy
under grant DE-FG02-00ER41132. BGS thanks the REU program of the University
of Washington and the National Science Foundation for support.

\end{document}